\documentclass[twocolumn]{aastex63}

\usepackage{amsmath}
\usepackage{txfonts}
\usepackage{amssymb}
\usepackage{graphicx}
\usepackage{epsfig}
\usepackage{color}
\usepackage{natbib}

\newcommand{\bl}[1]{\mbox{\boldmath$ #1 $}}
\def\gtaprx {\lower .1ex\hbox{\rlap{\raise .6ex\hbox{\hskip .3ex
	{\ifmmode{\scriptscriptstyle >}\else
		{$\scriptscriptstyle >$}\fi}}}
	\kern -.4ex{\ifmmode{\scriptscriptstyle \sim}\else
		{$\scriptscriptstyle\sim$}\fi}}}
\def\ltaprx {\lower .1ex\hbox{\rlap{\raise .6ex\hbox{\hskip .3ex
	{\ifmmode{\scriptscriptstyle <}\else
		{$\scriptscriptstyle <$}\fi}}}
	\kern -.4ex{\ifmmode{\scriptscriptstyle \sim}\else
		{$\scriptscriptstyle\sim$}\fi}}}

\newcommand{\cutt}[1]{\textcolor{blue}{}}

\newcommand{\Ms}{{\ensuremath{{M}_{\odot}} }}
\newcommand{\Ls}{{\ensuremath{L_{\odot} }}}
\newcommand{\Zs}{\ensuremath{Z_\odot}}
\newcommand{\C}{{\ensuremath{^{12}\mathrm{C}}}}

\newcommand{\N}{{\ensuremath{^{14} \mathrm{N}}} }

\shorttitle{First Planets}
\shortauthors{Vorobyov et al.}

\begin{document}

\title{Planet Formation at Cosmic Dawn: Planetesimals in H$_2$O-Rich Disks Around Low-Mass Stars}

\correspondingauthor{Muhammad A. Latif}
\email{latifne@gmail.com}

\author{Eduard I. Vorobyov}

\affiliation{Institut fur Astro- und Teilchenphysik, Universitaat Innsbruck, Technikerstrasse 25, 6020 Innsbruck, Austria}

\affiliation{Research Institute of Physics, Southern Federal University, Rostov-on-Don 344090, Russia} 

\author{Daniel J. Whalen}

\affiliation{Institute of Cosmology and Gravitation, Portsmouth University, Dennis Sciama Building, Portsmouth PO1 3FX}

\author{Muhammad A. Latif}

\affiliation{Physics Department, College of Science, United Arab Emirates University, PO Box 15551, Al-Ain, UAE (latifne@gmail.com)}

\author{Alexander Skliarevskii}

\affiliation{Research Institute of Physics, Southern Federal University, Rostov-on-Don 344090, Russia} 

\author{Christopher Jessop}

\affiliation{Institute of Cosmology and Gravitation, Portsmouth University, Dennis Sciama Building, Portsmouth PO1 3FX}

\author{Ryoki Matsukoba}

\affiliation{National Institute of Technology, Kochi College, 200-1 Monobe, Nankoku, Kochi 783-8508, Japan}

\author{Takashi Hosokawa}

\affiliation{Department of Physics, Graduate School of Science, Kyoto University, Sakyo, Kyoto 606-8502, Japan}

\author{Devesh Nandal}

\affiliation{Center for Astrophysics, Harvard and Smithsonian, 60 Garden St, Cambridge, MA 02138, USA}

\begin{abstract}

Primordial, or Pop III, supernovae (SNe) were the first, great nucleosynthetic engines in the Universe, forging the heavy elements required for the later formation of planets, and life.  Past studies suggest that the rise of planet formation was gradual, and did not peak until about half of the present age of the Universe after cosmic mean metallicities exceeded a critical value.  However, Pop III pair-instability (PI) SNe, which can eject over 100 \Ms\  of metals, locally enriched gas to metallicities of up to 1 \Zs\ at Cosmic Dawn, just 100 Myr after the Big Bang.  Here we show that planetesimals, the precursors of terrestrial planets, can form around low-mass, long-lived stars in the debris of such explosions, before the first galaxies and far earlier than previously thought. We modeled the collapse of a dense core with a Jeans mass of just 1 - 2 \Ms\ from a PI SN remnant and found that a protoplanetary disk formed with several Earth masses of planetesimals 0.5 - 1.0~au from their parent star, within its water snow line. The disk has H$_2$O mass fractions that are only a factor of a few less than in the Solar System today, raising the possibility of enrichment of the first planets in the Universe with water in direct analogy to Earth in the Solar system.
  
\end{abstract}



\section{Introduction}

Pop III stars ended the cosmic Dark ages at $z \sim$ 25 and began to chemically enrich the Universe with the heavy elements required for the later formation of planets and life \citep{wet08a,jet09b,hw10}.  In large-scale numerical simulations this process is gradual, with cosmic metallicities rising slowly over time.  For this reason, previous studies found that the rise of planet formation was gradual, and did not peak until cosmic mean metallicities exceeded a critical value at about half of the present age of the Universe \citep{jl12,bp15,zack16}.  However, cosmological simulations in large boxes do not resolve local metal mixing or cooling in SN remnants or the formation of clumps from which protoplanetary systems can emerge on scales of $\lesssim$ 1000 au \citep[e.g.,][]{wet25a}.  

It is now known that pair-instability SNe \citep[PI SNe;][]{rs67,brk67} of 120 - 260 \Ms\ Pop III stars can eject over 100 \Ms\ of metals \citep{hw02,jw11,yosh22,xing23}, enriching local gas to metallicities of up to $\sim$ 1 \Zs\ at $z \sim$ 20 and producing dense cloud cores that can collapse and form new stars in only a few Myr \citep{latif20c}.  Less energetic core-collapse (CC) SNe that expel 0.1 - 0.3 \Ms\ of metals can enrich local gas to $\sim$ 10$^{-4}$ \Zs\ on timescales of 50 - 100 Myr \citep{brit15,cw19}.  New cosmological simulations show that oxygen from the explosion can produce H$_2$O mass fractions of up to 10$^{-4}$ and Jeans masses of only 1 - 2 \Ms\ in such cores, raising the possibility of low-mass, long-lived star formation in water-rich environments among the first generation of stars in the Universe \citep{wet25a}.  We have now performed simulations that reveal for the first time that water-rich protoplanetary disks with planetesimals, the precursors of rocky planets, could form at Cosmic Dawn, billions of years earlier than previously thought.  In Section 2 we briefly describe our protoplanetary disk simulations and discuss disk evolution and planetesimal growth in Section 3.  We conclude in Section 4 with the prospects for the formation of planets with water at $z \sim$ 20 and their detection in low-metallicity surveys today.  

\section{Numerical Model}

The formation of the dense metal-enriched core in the PI SN remnant at $z \sim$ 17 in cosmological environments is described in detail in \citet{wet25a}.  Here, we describe how we follow its collapse into a protoplanetary disk with FEOSAD \citep[Formation and Evolution Of a Star And its Disk;][]{VorobyovAkimkin2018,VorobyovElbakyan2023}, a two-dimensional ($r,\phi$) protoplanetary disk code  that has been optimized for low-metallicity environments \citep{met22}.  It solves the continuity, momentum and energy equations of hydrodynamics for the gas and dust disk subsystems in the thin-disk limit, in which motion is restricted to the disk midplane and vertically integrated disk hydrodynamic quantities are evolved.  The basic equations, dust growth scheme, and dust-to-planetesimal conversion method are described in Appendix~\ref{App:num}.

We approximate the inner 1.0 \Ms\ of the PI SN core, the Jeans mass of this enclosed gas, by surface density and angular velocity profiles that have a compact central plateau and further out are inversely proportional to radius, which is typical of vertically integrated, gravitationally contracting clouds \citep{DappBasu2009}.  The central density and radius of the model core are $\sim 10^8$~cm$^{-3}$ and 9200~au, respectively, and the radius of the central plateau is 200~au.  The particular form of these profiles has minor effect on the properties of the protostellar disk for a given mass \citep{2012ARep...56..179V}.  The initial ratio of rotational to gravitational energy in the core is taken to be $\beta=0.2\%$, which is lower than usually assumed for massive primordial cores, 1-5\% \citep{Riaz2023,Sharda2025}, but in line with simulations of metal-poor cores with metallicity $\ge 10^{-4}$ \Zs\ \citep[e.g.,][]{Chon2024}.  Smaller values of $\beta$ may result from efficient magnetic braking while larger values lead to the formation of binaries, which is the subject of a separate study.

Following the efficiency of metal to dust conversion in the Solar neighborhood, we assume that half the metals in the core deplete onto dust grains so we set the initial dust-to-gas ratio $\xi^{\rm init}_{\rm d2g} = 4\times10^{-4}$. This corresponds to a metallicity $Z =$ 0.08 \Zs, slightly higher than that in the core but on the low end of metallicities of similar dense clumps in PI SN remnants \citep{latif20c}. Dust-to-gas mass ratios in galaxies in the local Universe at similar metallicities appear to be on average lower by a factor of several \citep{Remy2014}, but with a large scatter that includes our values. The effect of stronger metal depletion and thus lower $\xi^{\rm init}_{\rm d2g}$ is considered in Section~\ref{Sect:paramspace}.  The initial distribution of radii of dust grains in the core varies from 0.005 - 1.0~$\mu$m and the slope of the dust size distribution is fixed at $p=3.5$ (see Appendix~\ref{Sect:dustgrowth}). Cosmological galaxy formation models usually begin with a log-normal distribution \citep{2026OJAp....959986N}, but dust shattering due to mutual collisions recovers a power law  distribution with $p\approx 3.3$ \citep{2018MNRAS.478.2851M}.  A similar process can operate in a high redshift protoplanetary disk when dust growth is halted and dust grains are shattered at the collisional barrier in the inner several tens of au. Initial species mass fractions and gas and dust temperatures were obtained from a one-zone model of prestellar clouds for the metallicity of the PI SN core \citep{om05}. 

The numerical simulations start from the gravitational collapse of the slowly rotating model core. When the gas density exceeds $\sim 10^{11}$~cm we introduce a point-mass gravitating star in the center of the collapsing cloud. The star accretes from a circumstellar disk, which forms as a result of angular momentum conservation when the inner infalling layers of the core reach the centrifugal barrier. The evolution gradually continues into the embedded phase where the accreting star is surrounded by a rotationally supported accretion disk along with the remaining infalling envelope. All materials from the envelope land on the outer edge of the disk under the thin-disk approximation, consistent with trajectory calculations of infalling matter in three-dimensional envelopes by \citet{Visser2009}.  We end the simulation after $\sim 0.1$~Myr, when about 50\% of the core has accreted on the star plus disk system. The numerical resolution is $400\times 256$ cells on the polar grid ($r,\phi$) with a logarithmically spaced grid in the radial direction. The numerical resolution near the inner computational boundary is $\sim 10^{-2}$~au and it decreases to $\approx 1.3$~au near the disk's outer edge ($r=50$~au).

\section{Results}
\label{Sect:results}

\begin{figure*}
\center
\begin{tabular}{c}
\includegraphics[scale=0.69]{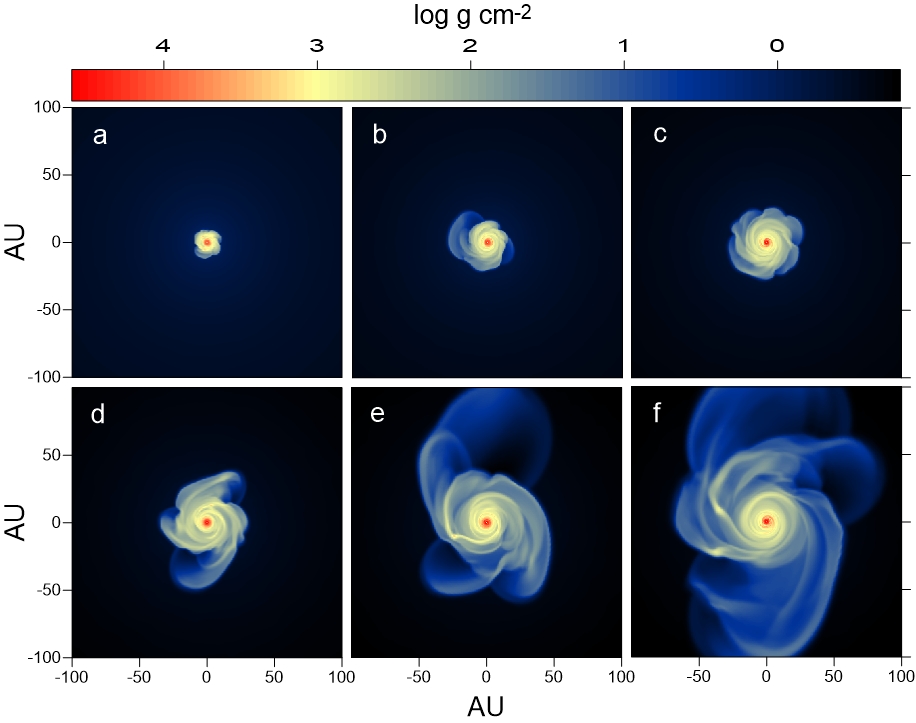}  \\
\includegraphics[scale=0.75]{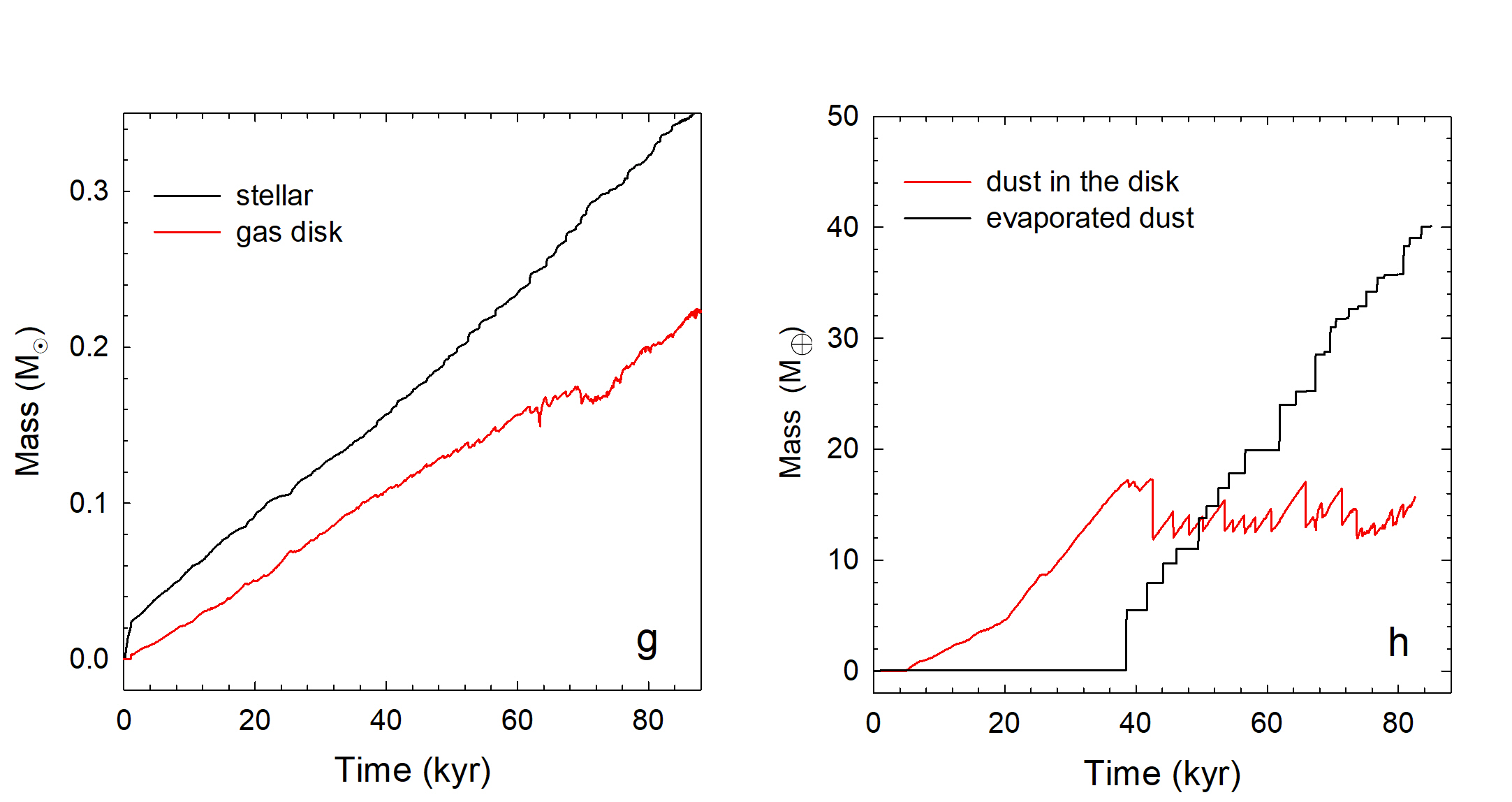} 
\end{tabular}
\caption{Formation and evolution of the disk.  Top:  gas surface densities (in log g cm$^{-2}$) for the inner 200~au of the disk are shown at 13, 21, 30, 40, 50 and 60 kyr after protostellar birth in (a) - (f).  Bottom:  star and disk gas masses after the birth of the protostar (g) and evaporated and remaining dust masses in the disk (h).}
\label{fig:diskevol}
\end{figure*}

We show the formation and evolution of the protoplanetary disk in Figure~\ref{fig:diskevol}. The protostar forms 24~kyr after the onset of collapse and the disk forms $\sim 1.0$~kyr later as spinning up and inspiralling material in the PI SN core reaches the centrifugal barrier near the protostar. In this early stage of evolution, infall rates exceed accretion rates onto the protostar, so the disk grows in mass and radius. By just 21~kyr after protostar formation, the disk develops a compact but clear spiral structure, a signature of the onset of the gravitational instability (GI).  As the disk size increases, the spiral pattern becomes more pronounced.

By the end of the simulation the star has reached 0.4~\Ms\  but its final mass will grow higher, given the initial core mass of $1.0~M_\odot$ and likelihood of additional mass accretion from immediate surroundings. The gas disk mass is about 50-60\% that of the star. The dust mass in the disk rises to $\sim$ 15~M$_\oplus$ owing to mass supply from the parent cloud but then saturates. It levels off because luminosity bursts brighter than a few tens of \Ls\ appear in the disk $40$~kyr after the formation of the protostar as explained below in Section~\ref{Sect:lumbursts} and evaporate the dust by heating it above the sublimation threshold. The evaporated dust mass increases with time, mirroring the occurrence of the bursts.  

\subsection{Onset of the GI}

\label{Sect:GI}

\begin{figure}
\centering
\includegraphics[scale=0.40]{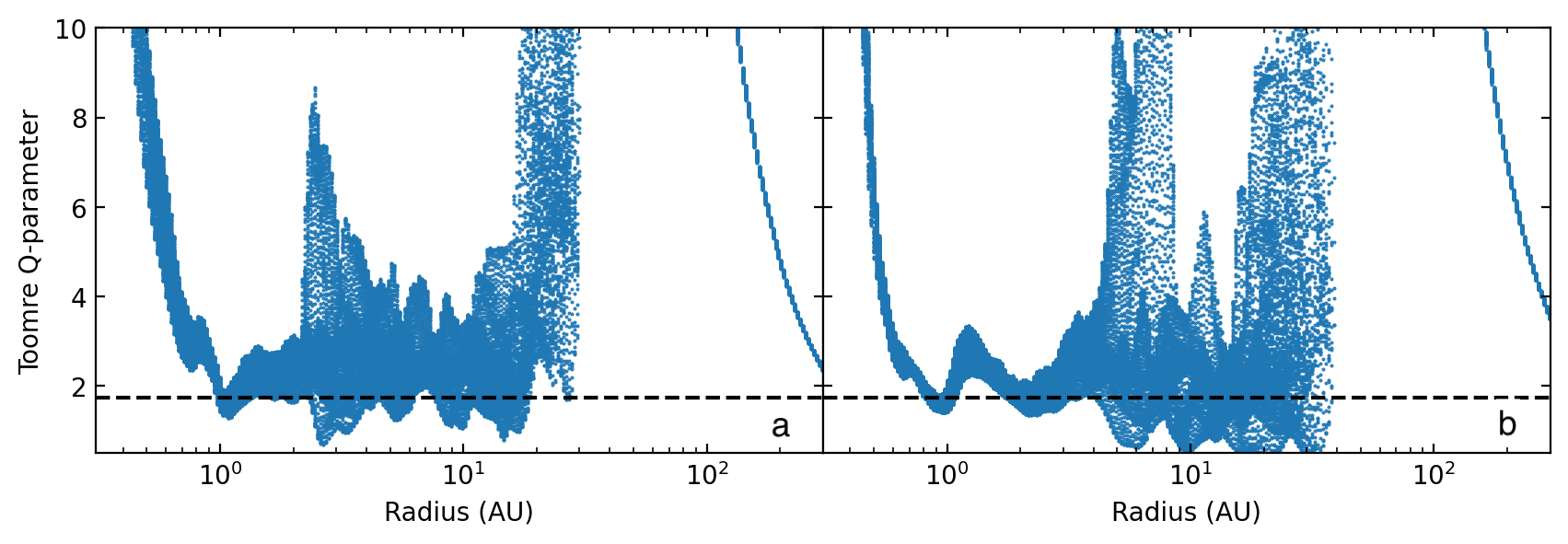} \\
\includegraphics[scale=0.43]{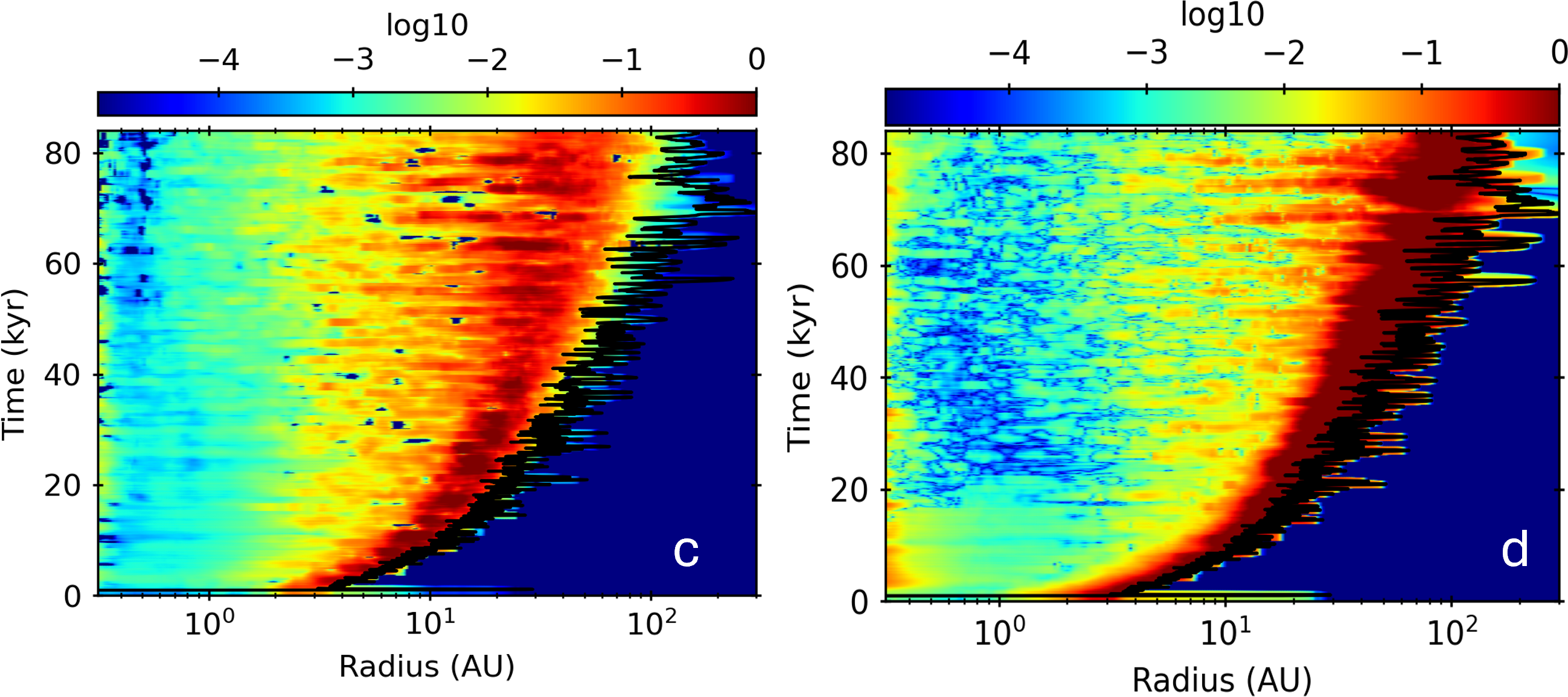} 
\caption{Top: radial profiles of the Toomre $Q$-parameter at 30 kyr (a) and 40 kyr (b) after protostellar birth. The horizontal dashed line marks the threshold value of $\sqrt{3}$ below which the gravitational instability develops.
Bottom:  effective $\alpha_{\rm GI}$ parameter (c) and Reynolds $\alpha_{\rm R}$ parameter (d) for the disk.  The values of $\alpha_{\rm GI}$ and $\alpha_{\rm R}$ are lowest in the innermost disk regions and highest near the disk outer edge, which is outlined by the black line. Note that $\alpha_{\rm GI}$ and $\alpha_{\rm R}$ are higher than $\alpha_{\rm visc}$ throughout most of the disk.
}
\label{fig:qpar}
\end{figure}

The GI transports mass radially inward via gravitational torques \citep{VB2007,Riols2017}. The susceptibility of the disk to the GI is shown in the top panels of Figure~\ref{fig:qpar} by the radial distribution of the Toomre $Q$-parameter at $t = $ 30~kyr and 40~kyr after protostar formation, when major episodes of planetesimal formation take place (see Section~\ref{Sect:paramspace}). The $Q$-parameter is defined as  
\begin{equation}
    \label{eq:ToomreQ}
    Q = \dfrac{\tilde{c}_{\rm s} \Omega} {\pi G \left( \Sigma_{\rm g} + \Sigma_{\rm d} \right)},
\end{equation}
where $\tilde{c}_{\rm s} = c_{\rm s} \sqrt{1 + \xi_{\rm d2g}}$ is the sound speed modified for the presence of dust \citep{VorobyovAkimkin2018}, $\xi_{\rm d2g}$ is the local dust-to-gas ratio, $\Sigma_{\rm g}$ and $\Sigma_{\rm d}$ are the surface densities of gas and dust (both components), $\Omega$ is the angular velocity of the gas, and $G$ is the gravitational constant. Since azimuthal averaging of the $Q$-parameter may be misleading, the $Q$-values in each cell along the azimuth (at a given distance $r$) are plotted. Regions of the disk from a few au to a few tens of au are prone to the GI because $Q$-values there systematically fall below the threshold value of $\sqrt{3}$ \citep{Polyachenko1997}.

To quantify the efficiency of mass transport by gravitational torques in protoplanetary disks, we calculate the effective $\alpha_{\rm GI}$ parameter: 
\begin{equation}
\alpha_{\rm GI} = {G_{r\phi} \over P \left| \frac{d \ln \Omega}{d \ln r} \right|},
\label{eq:alpha-GI}
\end{equation}
where $P$ is the gas pressure at the disk midplane and $G_{r \mathrm{\phi}}$ is the ($r,\phi$)-component of the gravitational stress tensor in the plane of the disk \citep{Riols2018},
\begin{equation}
\label{eq:grav_stress}
G_{r \mathrm{\phi}} = \frac{1}{4 \pi G r} \frac{\partial \Phi}{\partial r} \frac{\partial \Phi}{\partial \phi},
\end{equation}
where $\Phi$ is the gravitational potential in the disk, whose calculation is described in Appendix~\ref{App:num}. Using $\alpha_{\rm GI}$ to evaluate the efficiency of mass and angular momentum transport by gravitational torques is justified for our disk because the disk-to-star mass ratio is sufficiently high, $\ge 0.2$ \citep[see Figure~\ref{fig:diskevol} and][]{Vorobyov2010}.

As shown in the bottom left panel of Figure~\ref{fig:qpar}, $\alpha_{\rm GI}$ is smallest near the inner 1.0~au of the disk but larger in its outer regions, so mass transport is strongest in the intermediate and outer regions and weakest in the inner few au.  This gradient in $\alpha_{\rm GI}$ creates a bottleneck in the inner several au where mass transport is reduced and the GI is suppressed by high temperatures and strong shear \citep{Vorobyov2024}. Toomre $Q>>1$ in the inner disk have also been found in other global disk formation simulations \citep{Tomida2017,Durisen2023}.  

Reynolds stresses present another mechanism that can efficiently transport mass and angular momentum in protoplanetary disks, as was recently shown by, e.g., \citet{Commercon2026}.  Following their work, we calculated the Reynolds torque $T_{r\phi}^{\rm R}$ and the Reynolds $\alpha_{\rm R}$ as 
\begin{equation}
    T_{r\phi}^{\rm R} = \Sigma_{\rm g} \, \delta v_{r} \,\delta v_{\phi} ;  \,\,\,\, \alpha_{\rm R} = { \langle T_{r\phi}^{\rm R} \rangle \over \langle  {\cal P} \rangle \left| \frac{d \ln \Omega}{d \ln r} \right| }, 
\end{equation}
where the brackets denote azimuthal averaging, $\delta v_r$ and $\delta v_\phi$ are deviations of the instantaneous radial and azimuthal velocities of the gas $v_r$ and $v_\phi$ from the corresponding azimuthally averaged velocities $\langle v_r \rangle$ and $\langle v_\phi \rangle$,  and $\cal P$ is the vertically integrated gas pressure.  As shown in the bottom right panel of Figure~\ref{fig:qpar}, the trend in the radial distribution of $\alpha_{\rm R}$ follows that of $\alpha_{\rm GI}$, with smallest values in the inner several au and highest ones near the disk's outer edge where the velocity fluctuations due to spiral arms are strongest (see Figure~\ref{fig:diskevol}).  This behavior can be expected for gravito-turbulent disks with velocity fluctuations mainly generated by the GI.

\begin{figure}
\center
\begin{tabular}{c}
\includegraphics[scale=0.5]{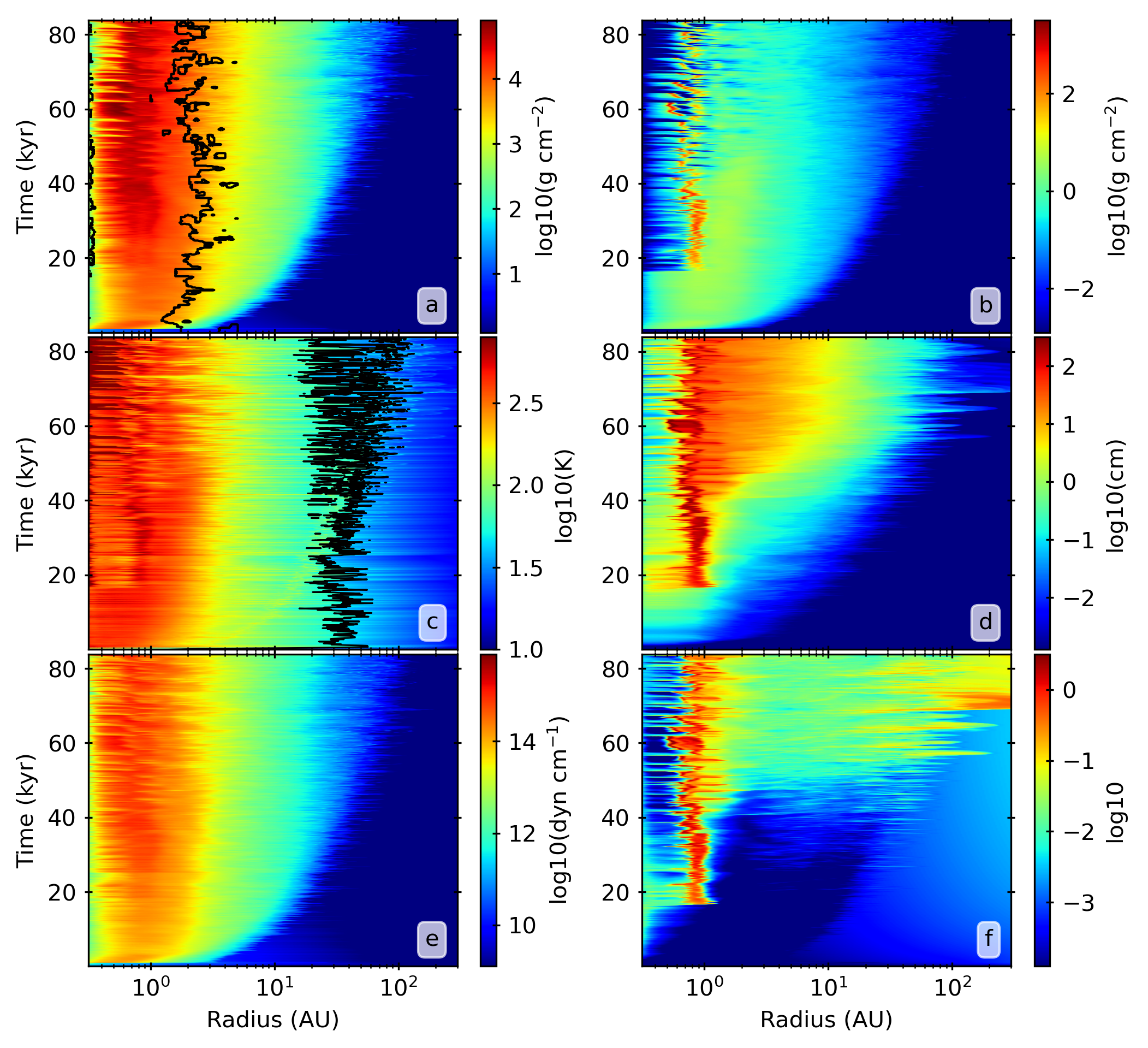} 
\end{tabular}
\caption{
Evolution of disk properties over time.  Azimuthally averaged gas surface densities, dust surface densities, gas temperatures, maximum dust grain radii, vertically integrated gas pressures, and Stokes numbers for the disk are shown in (a) - (f), respectively. The black line in panel (a) outlines the GI-suppressed dead zone in the inner disk as defined by $\alpha_{\rm GI} < 5\times 10^{-3}$.  The black line in panel (c) marks radii in the disk where temperatures fall below the CMB floor at $z =$ 17, about 49 K.}
\label{fig:alphaGI}
\end{figure}

A region with reduced mass and angular momentum transport forms in the inner disk, characterized by local gas density and pressure enhancements as shown in Figure~\ref{fig:alphaGI}.  This "dead zone" has characteristics similar to those in layered disk models \citep{Gammie1996}, except that the dead zone here is due to the suppression of the GI, not the suppression of magnetorotational instability (MRI, but see Section~\ref{Sect:paramspace}). A detailed comparison of MRI-suppressed and GI-suppressed dead zones was done in \citet{Vorobyov2024}. We note that $\alpha_{\rm GI}$ and $\alpha_{\rm R}$ are negligible prior to the formation of the disk ($t < 25$~kyr) and beyond the disk outer edge at later evolution times because the spatial distribution of matter there is nearly axisymmetric, the gravitational torques are small, and the fluctuations in the flow are minimal. Also,  $\alpha_{\rm GI}$  and $\alpha_{\rm R}$ exceed the background $\alpha_{\rm visc}=10^{-4}$ in our model (which is non-zero due to assumed low-level hydrodynamic turbulence) by several orders of magnitude  everywhere except in the innermost regions of the disk.  The $\alpha_{\rm GI}$ profile in our disk is similar to those in other recent simulations \citep{xu23}, which is an independent confirmation of the physics in our model.  The occurrence of the GI-suppressed dead zone is not due to simplified initial conditions in the cloud core.  Numerical models with and without turbulence in the core show the development of gravitational instability in the disk soon after its formation \citep{Tomida2017,Bate2018,Wurster2020}, which has the general property of being stronger in the outer part of the disk and weaker at smaller distances.


\subsection{Dust Accumulation / Streaming Instability}
\label{Sect:SI}

\begin{figure}
\centering
\includegraphics[scale=0.45]{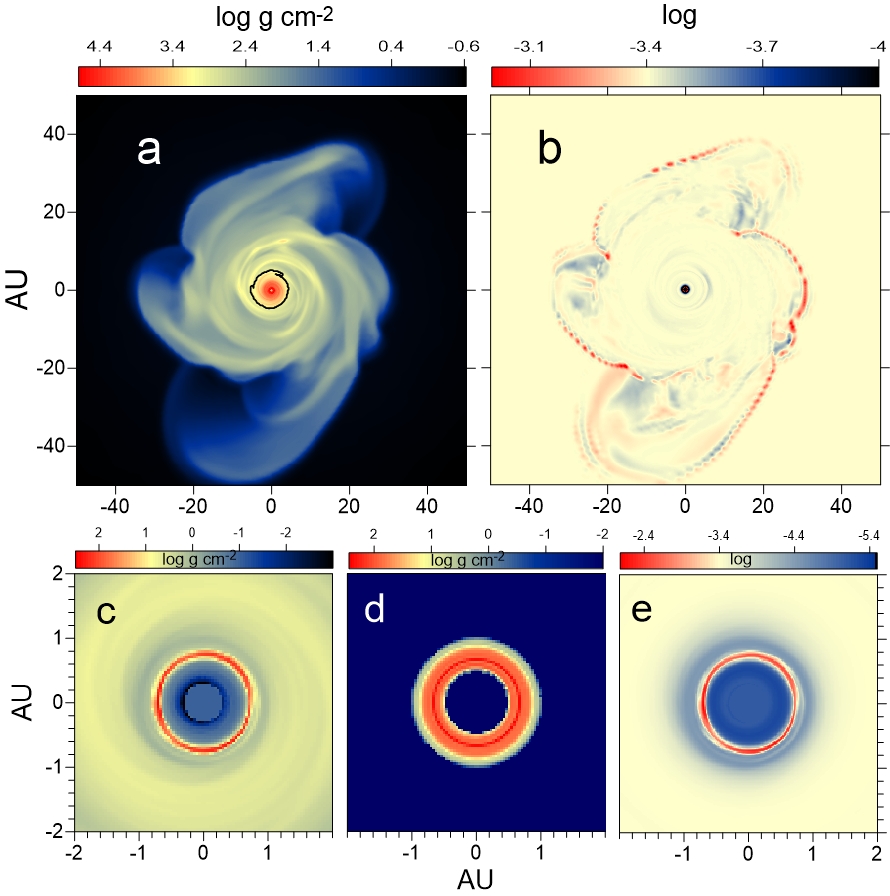} 
\caption{Gas, dust and planetesimal surface densities, alongside the dust-to-gas mass ratio $\xi_{d2g}$.  Gas densities and $\xi_{d2g}$ are shown 40 kyr after the formation of the protostar in (a) and (b), respectively, and (c)-(e) show the dust and planetesimal densities and $\xi_{d2g}$ in the central 4~au of the disk, respectively. The black line marks the water snowline in (a). The initial $\log \xi_{\rm d2g} = -3.4$ is shown in yellow. A movie of the gas and dust densities in the disk as a function of time, together with the dust-to-gas mass ratio normalized to the initial value, can be viewed in the supplementary online material.}
\label{fig:1FEOSAD}
\end{figure}

As shown in panel (a) of Figure~\ref{fig:alphaGI}, gas accumulates in the inner several au because of the bottleneck effect. Together with peak temperature (panel c), this creates the gas pressure maximum visible in panel (e).  Panels (d) and (f) show that dust in the bulk of the disk grows to radii greater than 1.0~mm and that its Stokes number exceeds $10^{-2}$, so it decouples from the gas and begins to efficiently drift inward. However, this inward drifting dust is effectively trapped in the GI-suppressed dead zone, just as dust is captured in the MRI-suppressed dead zones \citep{Flock2015,Pinilla2016}, leading to a local dust enhancement that in turn promotes dust growth to radii $\ge 1.0$~cm via mutual collisions.  Approximately 20~kyr after disk formation, the dust-to-gas mass ratio and the Stokes number exceed 0.01 and 0.1, respectively, so conditions in the inner disk become favorable to the development of the streaming instability (Equations~\ref{eq:SI_cond_1} and \ref{eq:SI_cond_2}), converting cm-radius dust grains into planetesimals, the main building blocks of planets.  

Figure~\ref{fig:1FEOSAD} shows the global structure of the disk and its inner several au where the dead zone is located and the process of dust to planetesimal conversion according to (Equation~\ref{eq:dpltsm}) takes place.  A narrow ring of dust with $\xi_{\rm d2g}$ as high as $6\times 10^{-3}$ is visible at $\sim$ 0.5~au, while the planetesimals are spread around the dust ring in an annulus with a width of $\sim$ 0.5~au. The ring of planetesimals is broader because the dust ring slightly migrated in and out as the disk evolved.  As shown in Figure~\ref{fig:alphaGI}, dust grains in the inner disk grow to radii much larger than 1.0~mm and Stokes numbers become greater than 0.01, thus overcoming the dust grain radius and Stokes number barriers for the development of the streaming instability in the dead zone \citep[][see also Appendix~\ref{App:plt}]{Carrera2022}. 

The black contour in panel (a) of Figure~\ref{fig:1FEOSAD} marks the water snowline in the disk midplane at 40 kyr (Appendix~\ref{App:snowline}).  The planetesimals in our model form interior to the snowline and thus are expected to be water deficient.  We note that the protoplanetary core is expected to be more abundant in water \citep{wet25a} than assumed above, in which case the radial position of the snowline can shift somewhat inwards. 

Dust is converted to planetesimals but their dynamics are not computed in this work as it would require combining FEOSAD with N-body codes. The subsequent conversion of planetesimals to planetary cores may depend on the subtleties of planetesimal dynamics and their mass spectrum. Nevertheless, we show in the next section that the total mass of planetesimals in our disk is sufficient to form a rocky planet with mass similar to that of Mars or Earth. This can occur via oligarchic growth of colliding planetesimals \citep{2000KokuboIda,2006Icar..180..496C}, especially if the growth process of the most massive "oligarchs" is promoted with accretion of pebbles \citep{2014LambrechtsJohansen}, which form abundantly in our model as shown in Appendix~\ref{App:pebbles}.  

\subsection{Luminosity Bursts}

\label{Sect:lumbursts}

\begin{figure}
\centering
\begin{tabular}{c}
\includegraphics[scale=0.52]{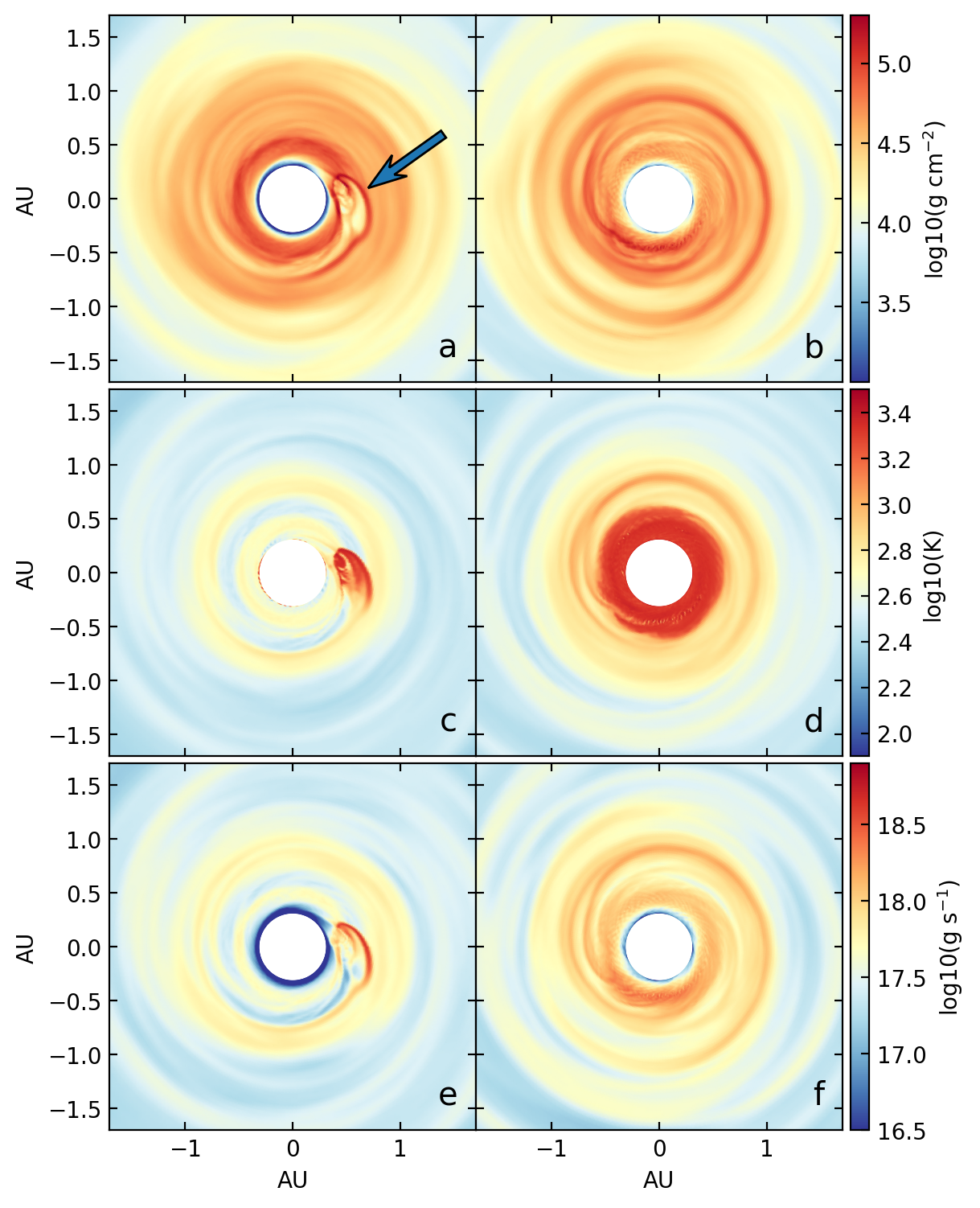}  \\
\includegraphics[scale=0.5]{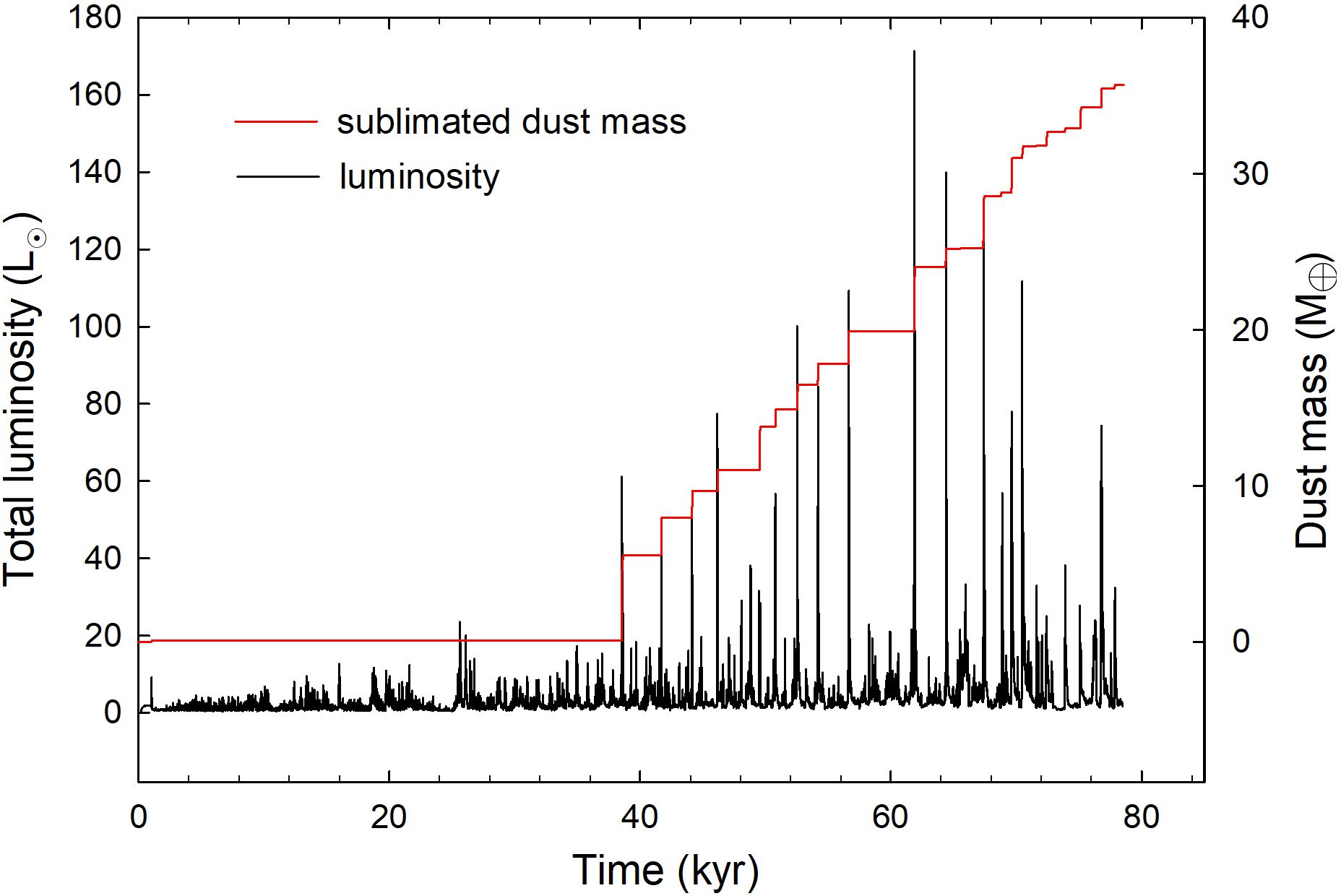} 
\end{tabular}
\caption{Top: the origin of luminosity bursts.  The central 3 au of the disk just before (a, c and e) and during (b, d, and f) a protostellar accretion burst. Rows from top to bottom show gas surface densities, gas temperatures, and turbulent dynamical viscosities.  The arrow indicates a local density disturbance that produces viscous heating in the inner disk.  Bottom:  total luminosities (accretion plus photospheric) and evaporated dust masses as a function of time from protostellar birth.}
\label{fig:burst}
\end{figure}


As mentioned in Section~\ref{Sect:results}, dust evaporation is correlated with luminosity bursts that develop in the disk 40 kyr after protostellar birth.  They are caused by episodic increases in protostellar accretion rate, $\dot{M}$, like those routinely observed in the Milky Way \citep{2014prpl.conf..387A,2023ASPC..534..355F} but they have also been predicted to occur during low-metallicity protoplanetary disk formation \citep{kad21, 2021MNRAS.500.4126M,Klessen2023}.  Outbursts in protostars today, known as FU~Orionis-type objects, are thought to be caused by various disk instabilities, planet-disk interactions or close stellar flybys \citep[see][for review]{2014prpl.conf..387A}.  Here, the bursts originate from strong, spiral-like gas density perturbations in the innermost disk like the one visible in panel (a) on the left in Figure~\ref{fig:burst} that are due to brief episodes of the gravitational instability.  These episodes are accompanied by an increase in turbulent viscosity and local heating, as shown in panels (c) and (e). After a few orbital periods the perturbation is spread over the entire azimuth, warming the innermost disk to above 1000~K as seen in panel (d).  As a result, $\dot{M}$ rises sharply by factors of 10-100. This burst mechanism is unique to low-metallicity disks and will receive more attention in follow-up studies.

The gravitational energy of this accreted matter is transformed into heat and radiated away as accretion luminosity, $L_{\rm acc}=0.5 G M_\ast \dot{M} / R_\ast$, where $M_\ast$ and $R_\ast$ are the mass and radius of the protostar. The accretion luminosity during the bursts greatly exceeds that of the photosphere, which is powered by deuterium burning and compressional heating.   As shown in the bottom plot in  Figure~\ref{fig:burst}, luminosity bursts regulate the conversion of dust to planetesimals by heating the inner disk and evaporating dust.  The sublimated dust mass exceeds that of the dust remaining in the disk after $\sim$ 50~kyr of evolution. Some of the dust vapor is accreted onto the protostar with the gas but some may return as sub-$\mu$m dust seeds as the disk cools after the burst, but they are not included in our simulation.  

\subsection{Total Planetesimal Mass}

\label{Sect:paramspace}
\label{Sect:tot-plt-mass}

\begin{figure}
\centering
\begin{tabular}{c}
\includegraphics[scale=0.7]{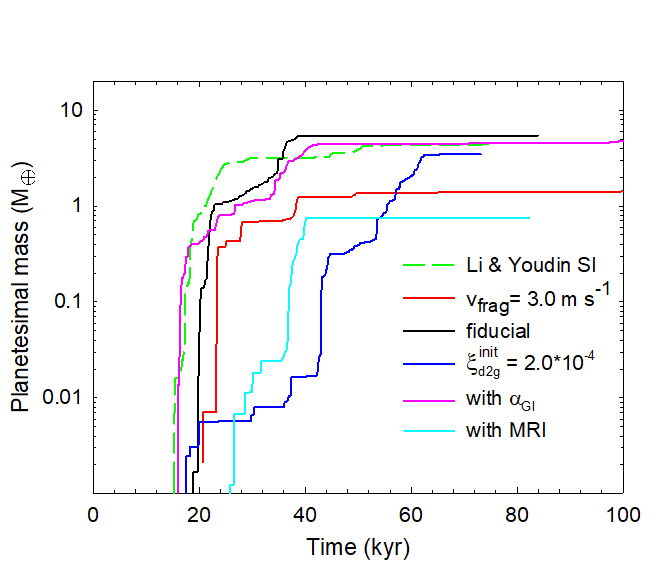} 
\end{tabular}
\caption{Total planetesimal masses for the fiducial model (solid black) and parameter study models. The latter include the models with $v_{\rm frag}=$3.0~m~s$^{-1}$ (solid red), $\xi^{\rm init}_{\rm d2g}=2\times10^{-4}$ (solid blue), with the streaming instability criteria from \citet{LiYoudin2021} (dashed green), MRI with solar metallicity excitation criteria (cyan), and gravitoturbulence (pink).}
\label{fig:parspace}
\end{figure}

We show total planetesimal masses, $M_{\rm plt}$, for our fiducial model with $v_{\rm frag}=5.0$~m~s$^{-1}$ and $\xi^{\rm init}_{\rm d2g}=4\times 10^{-4}$ in Figure~\ref{fig:parspace}. $M_{\rm plt}$ is found by integrating the rate of planetesimal formation, Equation~(\ref{eq:dpltsm}), over time and disk extent.  $M_{\rm plt}$ reaches $\sim$ 6 $M_{\oplus}$ in our fiducial run before leveling off at $t\approx 37$~kyr due to dust evaporation by luminosity bursts. This short epoch of planetesimal formation implies that their material is only weakly affected by long-term chemothermal processes.  The streaming instability depends on the dust size distribution and the maximum dust size, which is set by the disk conditions in our model (Section~\ref{Sect:dustgrowth}) and may also depend on several free parameters, such as the fragmentation velocity and the rate of turbulence. Assuming a lower $v_{\rm frag}=3.0$~m~s$^{-1}$ reduces final planetesimal mass to $\sim$ 1.5 $M_{\oplus}$ while using the more recent streaming instability criteria of \citet{LiYoudin2021} reduces $M_{\rm plt}$ to $\sim$ 5 $M_{\oplus}$.   We also consider the possibility that turbulent $\alpha_{\rm visc}$ is modified for the presence of GI, which can generate significant gravitoturbulence \citep{Riols2017}, so that the effective $\alpha$-parameter becomes the sum of $\alpha_{\rm visc}$ and $\alpha_{\rm GI}$ (see Figure~\ref{fig:alphaGI}), but this only slightly reduces the total planetesimal mass.  Lowering the initial dust-to-gas mass ratio $\xi_{\rm d2g}^{\rm init}$ to $2\times 10^{-4}$ (stronger metal depletion) suppresses planetesimal formation at $t<40$~kyr because luminosity bursts begin at earlier times but it resumes after the bursts subside, forming $3.5~M_\oplus$ by 80~kyr, which is only a factor of 1.7 lower than in the fiducial model with $\xi_{\rm d2g}^{\rm init}=4\times 10^{-4}$.  Additional models testing the boundary conditions and gravitational potential solver are discussed in Appendix~\ref{App:BCs}. 

Finally, we note that spiral arms at these early stages of disk evolution may be not be efficient sites of dust trapping \citep{Vorobyov2024}, unlike at later times when the effect can be strong \citep{Rice2025}. High densities and temperatures in young disks effectively lower the Stokes number of dust grains below 0.1 in the vicinity of spiral arms (see Figure~\ref{fig:alphaGI}), making dust trapping efficient only around the corotation radius \citep{VorobyovAkimkin2018}.  The lack of dust spirals in young embedded disks found by the eDisks survey seems to corroborate this point \citep{Ohashi2023}.  Since numerical constraints required us to resolve the inner regions of the disk at the expense of the outer ones, our conclusions may change when higher resolution can be applied at greater radii or the disk can be evolved to later times.  Also, the conversion of dust into planetesimals in spiral arms at the expense of the inner regions has only been shown to happen in disks at solar metallicities and for as long as the particle size distribution extends to $\mathrm{St}\sim 1.0$ \citep{Rice2025,Baehr2025}.  We note that a few clumps, which can be sites of efficient planetesimal formation, appear in spiral arms (Figure~\ref{fig:diskevol}) but they are quickly torn apart by tidal interactions with the spirals \citep{ZhuHartmann2012}.  Had they survived, they might have produced sub-solar mass companions, but planetesimal formation around the primary already occurs by the disk fragmentation epoch. The companion scenario merits further study with higher numerical resolution.

\subsection{Pebbles}

\label{App:pebbles}

\begin{figure}
\centering
\includegraphics[scale=0.65]{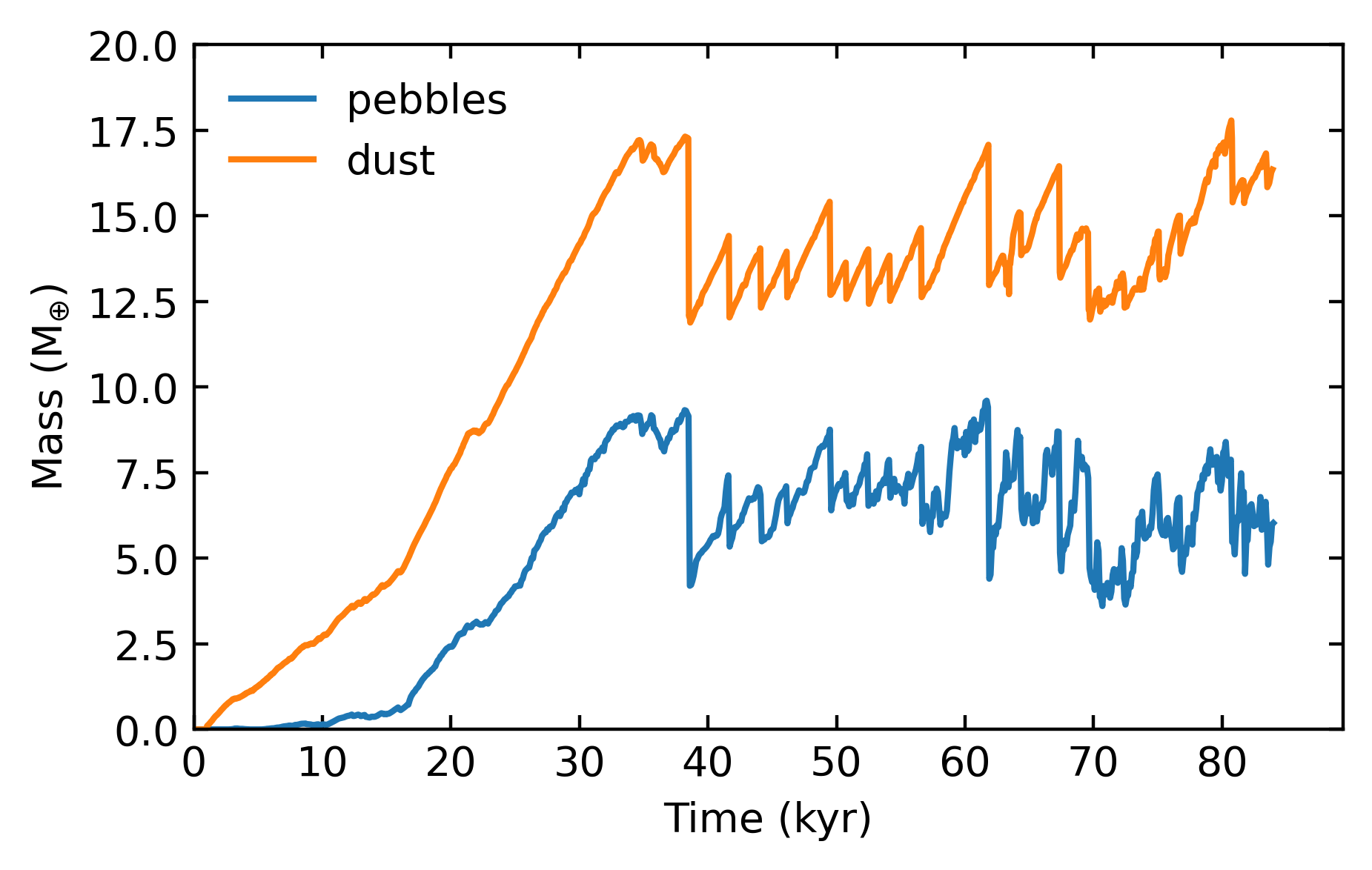} 
\caption{Dust and pebble masses in the disk as a function of time in the fiducial model.}
\label{fig:pebbles}
\end{figure}

We can estimate the mass reservoir of pebbles in the disk in the same manner as \citet{Topchieva2024} by first assuming that pebbles are dust particles with radii $a_{\rm peb,0} \ge$ 0.5~mm.  This lower limit is motivated by the typical sizes of chondrules, 0.1 - 1.0~mm \citep{2019ApJ...887..230M}.  Second, we stipulate that pebbles are larger than dust grains with radius $a_{\rm St_0}$ that have $\mathrm{St_0}=0.01$ locally in the disk, meaning that they can decouple dynamically from gas and be accreted by a protoplanetary core. These two conditions define the minimum radius of pebbles to be
\begin{equation}
a_{\rm peb,min}=
\begin{cases}
a_{\rm St_{\rm 0}}, & \text{ if } a_{\rm St_{\rm 0}} > a_{\rm peb,0}, \\
a_{\rm peb,0}, & \text{ if } a_{\rm St_{\rm 0}} \leq a_{\rm peb,0}.
\end{cases}
\label{eq:apebmin}
\end{equation}
Once $a_{\rm peb,min}$ is constrained, the mass of pebbles is found in each computational cell by calculating the mass of dust in the [$a_{\rm peb,min}:a_{\rm max}$] bin for the assumed slope of dust size distribution, $p=3.5$.  Previous studies have examined the effects of variations in $\mathrm{St_0}=0.01$ and $a_{\rm peb,0}=0.5$~mm \citep{Topchieva2024}. As shown in Figure~\ref{fig:pebbles}, the protoplanetary disk has on average several M$_{\oplus}$ of pebbles, which increases the likelihood of pebble-assisted formation of planetary cores in the debris of the PI SN explosion \citep[see, e.g.,][]{Lau2022}.

\subsection{Snowline}

\label{App:snowline}

The water snowline in the disk midplane is calculated by finding the radial position at which the partial and saturated water pressures are equal \citep{Okuzumi2016}. The partial water pressure is calculated assuming a constant water mass fraction of $10^{-4}$, as found for the dense core by \citet{wet25a}. 
We note that this fraction may change locally due to (i) radial transport of water ice on the surface of dust grains and (ii) chemical reactions involving cosmic rays and ultraviolet radiation. The former is a more dominant mechanism than chemical reactions in solar metallicity disks \citep{Booth2019}, while the latter may affect the water abundance in the upper rarefied layers of the disk \citep{2025AJ....170...67A}, but only on timescales $> 1.0$~Myr \citep[][although cosmic ray fluxes among the first generation of stars were small, as discussed in Section~\ref{sect:MRI}]{Eistrup2018}.  Future simulations dedicated to both mechanisms will shed light on their relative importance in low-metallicity disks.  The saturated pressure of water vapor is approximated by the Arrhenius form,
\begin{equation}
P_{\rm sat} =     \exp\left( - {L_{\rm H2O} \over T} + A_{\rm H2O} \right) \, \mathrm{dyn~cm^{-2}},
\end{equation}
where $T$ is the gas temperature, $L_{\rm H2O}=6070$~K is the heat of sublimation and $A_{\rm H2O}=30.86$ is a dimensionless constant.  We show the snowline in the disk at 40 kyr in panel (a) of Figure~\ref{fig:1FEOSAD}, where it can be seen to enclose the planetesimals.

\subsection{Magnetorotational Instability (MRI)}

\label{sect:MRI}

The MRI can drive turbulence that destroys dead zones and prevents planetesimal formation in magnetically active disks \citep{Kadam2022}. Magnetic fields could be present in our disk, even in the absence of metals. Turbulent subgrid dynamos in principle can build up disordered magnetic fields that are later ordered and amplified by rotation in primordial disks \citep{schob12,ls15,sharda20}.  However, the degree to which those field lines would be coupled to gas in the disk and ramp up turbulence via the MRI is much less certain, for several reasons.  Ohmic resistivity and ambipolar diffusion reduce the coupling of magnetic fields to gas and dampens the MRI \citep{Gressel2015}.  

Thermal ionization fractions in our disk due to alkali elements would be 13 times lower than those in disks at solar metallicity because it is at 0.08 \Zs, and smaller ionization fractions exacerbate Ohmic resistivity and ambipolar diffusion, further dampening the MRI.  The odd-even effect in PI SNe, in which even-numbered elements are preferentially synthesized over odd-numbered elements during the explosion \citep{hw02}, suppresses the abundances of alkali elements in the disk by another two orders of magnitude, even further reducing thermal ionization fractions and weakening the MRI.  Finally, there are few sources of non-thermal ionization such as cosmic rays because SNe were sparse in the primordial Universe. The PI SN remnant in which the disk formed likely produced some but with fluxes that were certainly lower than those in the Galaxy today.  These points are corroborated by recent work that found the frequency of MRI bursts in magnetically layered disks to be greatly reduced at low metallicities \citep{kad21}.  

We ran a model that approximated the effect of the MRI by setting $\alpha_{\rm visc} =$ 10$^{-2}$ where temperatures in the disk exceed 1300~K \citep{BaeZhu2014} and thermal ionizations become important and $\alpha_{\rm visc} =$ 10$^{-4}$ everywhere else.  As shown in Figure~\ref{fig:parspace}, we found that $\sim$ 0.7 $M_\oplus$ of planetesimals still form in the disk so the MRI reduces but does not prevent planetesimal formation.  Given that our approximation to the MRI assumes solar metallicity by not taking into account the much lower potassium fractions in the PI SN, it is truly a maximum upper limit to its impact on planetesimal masses here.  

\subsection{Cosmic Microwave Background (CMB)}

\label{sect:CMB}

Although we assumed a temperature $T =$ 10 K for the cloud core, the cosmic microwave background (CMB) would in reality impose a temperature floor $T_{\rm CMB} = 2.73 (1+z)$ K ($\sim$ 49 K at $z =$ 17) on it because CMB photons would heat it to this temperature if metals or dust cool it below this value.  However, as shown in panel (c) of Figure~\ref{fig:alphaGI}, gas in the disk less than 20 - 30 au from the central star rarely falls to the CMB floor because the low metallicity of our disk prevents efficient cooling.  Consequently, the CMB has little effect on the dynamics of the disk or planetesimal formation in our simulations although the true infall rates onto the disk may have been somewhat higher at early times.  

\section{Conclusion}

Local chemical enrichment by Pop III SNe likely led to planet formation at Cosmic Dawn, before the first galaxies and far earlier than previously thought.  In the PI SN cloud core we study here, the main prerequisites for planet formation at $z \sim$ 17 are fulfilled: dust growth to radii $\ge 1.0$~cm, dust enhancement above the dust-to-gas mass ratio of 1:100, onset of the streaming instability, and conversion of dust to planetesimals.  \citet{wet25a} considered the simplest case of a single massive star in a primordial halo so there were no other Pop III stars in the vicinity of the protoplanetary disk when it formed, and no other halos with stars within a kpc of the host halo.  Had other stars formed in the halo, they would have been several hundred pc from the protoplanetary disk because it formed in the expanding PI SN blast front.  Their ionizing UV flux would likely have had only a minor effect on the disk and its water mass fractions because the intervening gas was heavily enriched by metals and dust that would have attenuated the photons.   Also, because the dense cloud core forms in hydrodynamical instabilities in the outgoing blast front, the reverse shock in the PI SN remnant does not destroy dust or molecules in the disk because it never reaches it.

The low metallicity of the parent star could flag it as a potential primordial planetary system but by itself would not be conclusive because similar metallicities would arise in early galaxies at later times.  However, if the telltale 'odd-even' nucleosynthetic imprint of a PI SN were found in the star it would make a compelling case for the disk to have formed at Cosmic Dawn \citep{hw02}.  Furthermore, because luminosity bursts terminate planetesimal formation in the first several tens of kyr they should be composed of material that is mostly unprocessed by disk chemothermal processes and may, e. g., have peculiar isotopic compositions.  Planets forming from these planetesimals could be detected as worlds around ancient metal-poor stars in the Galactic halo in future exoplanet surveys such as PLATO \citep{bvd24}, since 0.4 \Ms\ stars from that era would still be on the main sequence today.   Although the protoplanetary disk in our simulation forms at $z \sim$ 17, others like it could have formed at even earlier times.  Dynamical heating \citep{wise19} and supersonic baryon streaming motions \citep{stacy11a,srg17} can produce cosmological halos that partially contain the PI SN at $z \sim$ 20 - 25 so water-rich protoplanetary disks with planetesimals might have formed only 50 - 100 Myr after the big bang.

As shown in Figure~\ref{fig:1FEOSAD}, the water snowline lies at several au from the protostar, with radial variations due to luminosity bursts.  Since the planetesimals in our model form within the snowline, they are water deficient.  However, with water mass fractions of $\sim$ 10$^{-4}$ in the disk \citep[only a factor of a few less than in the Solar System today;][]{wet25a}, planetary cores forming from the water-deficient planetesimals could later be enriched by water via accretion of asteroids from beyond the snowline that were created by other means \citep{Andama2024,Linn2025}, in direct analogy to how the Earth was enriched by water at early times \citep{2017SSRv..212..743P,2013Natur.499..328W}.  The possibility that rocky worlds with oceans may have formed among the first generation of stars in the Universe upends decades of thought on the cosmic origins of life.

\acknowledgments

We are grateful for constructive comments by the referee that improved the quality of our paper.  MAL was supported by UAEU UPAR grant No. 31S390.  CJ was supported by STFC grants ST/S505651/1 and ST/T506345/1.  RM and TH acknowledge support by Grants-in-Aid for Scientific Research (TH:19H01934, 19KK0353, 21H00041, 22H00149) from the Japan Society for the Promotion of Science.   DN was supported by the Swiss National Science Fund (SNSF) Postdoctoral Fellowship, grant number: P500-2235464.  The FEOSAD simulations were carried out on the Austrian Scientific Cluster (ASC).

\bibliographystyle{apj}
\bibliography{refs}

\appendix

\section{Numerical Method} 

\label{App:num}

FEOSAD evolves the gaseous component of the disk with the continuity, momentum and energy equations in the thin-disk approximation: 
\begin{equation}
\label{eq:cont}
\frac{{\partial \Sigma_{\rm g} }}{{\partial t}}   + \nabla  \cdot 
\left( \Sigma_{\rm g} {\bl v} \right) = 0,  
\end{equation}
\begin{equation}
\label{eq:mom}
\frac{\partial}{\partial t} \left( \Sigma_{\rm g} {\bl v} \right)  + \nabla \cdot \left( \Sigma_{\rm
g} {\bl v}\otimes {\bl v} \right)  =  - \nabla {\cal P}  + \Sigma_{\rm g} \, {\bl g}
+ \nabla \cdot \mathbf{\Pi}  - \Sigma_{\rm d,gr} {\bl f},
\end{equation}
\begin{equation}
\frac{\partial e}{\partial t} +\nabla \cdot \left( e {\bl v} \right) = -{\cal P} 
\left(\nabla \cdot {\bl v}\right) -\Lambda  + 
\left(\nabla {\bl v}\right):\Pi, 
\label{eq:energ}
\end{equation}
where  $\Sigma_{\rm g}$ is the gas surface density, ${\bl v}=v_r\hat{{\bl r}}+v_\phi \hat{{\bl \phi}}$ is the gas velocity in the disk plane, $\cal{P}$ is the vertically integrated gas pressure, ${\bl f}$ is the drag force per unit mass between gas and dust, $e$ is the internal energy per surface area, and $\Lambda$ is the sum of the cooling and heating terms pertinent to protoplanetary disks at low metallicity.  The first and last right-hand-side terms in Equation~(\ref{eq:energ}) are compressional and viscous heating, respectively.  

The gravitational acceleration in the disk plane ${\bl g}$ accounts for gas and dust self-gravity and the gravity of the central star after it is formed. The gravitational potential of the gas and dust components is calculated with the convolution method \citep{Binney1987,Vorobyov2024}. This method does not require explicit smoothing-lengths, which are often employed to avoid the singularity in the gravitational potential calculations \citep[e.g.,][]{Baruteau2008}, but has an implicit smoother proportional to the size of the grid cell. The explicit smoothing-length method can achieve a slightly better accuracy in computing the gravitational potential if the proper choice of the smoothing factor is applied but our approach is more stable because there is no universal formula to find the best-fit smoothing length. Validation tests and advantages and disadvantages of our method for the gravitational potential are discussed in the Appendix of \citet{Vorobyov2024}, and the final planetesimal masses produced by both methods are compared in Figure~\ref{fig:parspace}.

To compute the viscous stress tensor ${\bl \Pi}$, we parameterize the kinematic viscosity with the usual $\alpha$-parameter approach with $\alpha = 10^{-4}$ throughout the entire domain. Low-level turbulence can be driven by (magneto-)hydrodynamic instabilities that are unresolved in our global disk simulations and it supplies dust grains with the velocity dispersion needed for grain-to-grain collisions and dust growth. We note that $\alpha_{\rm visc}$ should be distinguished from the $\alpha$ due to GI, $\alpha_{\rm GI}$, which is introduced in Section~\ref{Sect:GI} to gauge the efficiency of mass transport by spiral density waves in the disk.  As shown in \citet{VorobyovBasu2009}, the dynamics of gravitationally unstable disks for such low $\alpha$ values is almost totally dominated by disk self-gravity and associated gravitational torques. Our code also uses artificial viscosity to smooth shock fronts over two grid cells, but its effect on disk evolution is negligible compared to turbulent viscosity caused by gravitational instability, as shown in \citet{VB2007}.

\subsection{Chemistry and Cooling/Heating}
The term $\Lambda$ in Eq.~(\ref{eq:energ}) accounts for the cooling and heating processes in low-metallicity disks
\begin{equation}
       \Lambda = \left(Q_{\rm cont}+Q_{\rm mole} + Q_{\rm chem} +Q_{\rm metal}\right) 2 H_{\rm g},
    \label{eq:coolQ}
\end{equation}
where $Q_{\mathrm{cont}}$ is the rate of energy change due to (i) radiative continuum cooling of gas and dust from the disk surface, (ii) collisional transfer of energy between gas and dust, (iii) disk heating by stellar and background irradiation, $Q_\mathrm{mole}$ is the H$_{2}$ and HD line cooling rates, $Q_{\mathrm{chem}}$ is the chemical cooling--heating rate (through chemical reactions), and $Q_{\mathrm{metal}}$ is cooling by fine-structure line emission of O I (63 $\mu$m) and C II (158 $\mu$m).   
Heating by stellar irradiation is calculated taking the disk flaring into account as described in \citet{VorobyovBasu2009}. The incidence angle of stellar radiation is computed using the radial variation of the gas scale height $H_{\rm g}$, which is calculated based on the assumption of local hydrostatic equilibrium in the gravitational field of the central star and disk. The disk aspect ratio $H_{\rm g}/r$ lies in the 0.05-0.2 limits, justifying the thin-disk assumption. All constituents of $Q_{\rm tot}$ are volumetric cooling or heating rates, a factor of $2H_{\rm}$ converts volumetric rates into vertically integrated quantities.

The net rate of internal energy change due to radiation transport is
\begin{equation}
    Q_{\rm cont} = 4 \pi \left[ \mu_{\rm em} - (\kappa_{\rm P,g} \rho_{\rm g}+\kappa_{\rm P,d} \rho_{\rm d}) J \right],
    \label{eq:contQ}
\end{equation}
where $\kappa_{\rm P,g}$ and $\kappa_{\rm P,d}$ are the mean Plank opacities of gas and dust. Their values are taken from \citet{Semenov2003} but scaled down to the low metallicities considered here, $Z=0.08~Z_\odot$ (fiducial model) and $Z=0.04~Z_\odot$ (one of the parameter study models, see Section~\ref{Sect:tot-plt-mass}).  A uniform scaling factor is applied to the gas opacities but we use the local dust-to-gas mass ratio to scale down the dust opacities, an approach that allows us to consider the effects of local deviations in the dust-to-gas mass ratio on the dust optical depth.  Dust growth lowers the dust opacities at temperatures 100-1000~K \citep{OpTool2021}, typical of the inner disk regions, which may favor the formation of planetesimals because of less efficient dust sublimation (Section~\ref{Sect:dustgrowth}), fewer MRI bursts (Section~\ref{sect:MRI}), and a more efficient SI (Section~\ref{Sect:SI}).

The emission coefficient is
\begin{equation}
    \mu_{\rm em} = {\sigma \over \pi} \left( \kappa_{\rm P,g} \,\rho_{\rm g} T_{\rm g}^4 + \kappa_{\rm P,d} \,\rho_{\rm d} T_{\rm d}^4 \right),
    \label{eq:mu-em}
\end{equation}
where $T_{\rm g}$ and $T_{\rm d}$ are the gas and dust temperatures. The gas and dust volumetric densities  $\rho_{\rm g}$ and $\rho_{\rm d}$ are calculated from their surface density counterparts and the corresponding vertical scale heights and $ T_{\rm g} = e \mu (\gamma-1) / (\Sigma_{\rm g} R)$, where the time- and space-variable mean molecular weight $\mu$ and ratio of specific heats $\gamma$  account for non-ideal effects.  The dust temperature is found from the energy balance on dust grains due to thermal emission, absorption of stellar ultraviolet radiation by dust grains, and collisional exchange of energy between dust grains and gas \citep{Omukai2010}
\begin{equation}
    \kappa_{\rm P, d} B(T_d) = \kappa_{\rm P,d} J - \Gamma_{\rm col},
    \label{eq:coll}
\end{equation}
where $\Gamma_{\rm col}$  is the heating rate of dust through collisions with gas particles \citep{Hollenbach1979}.
 Our method, therefore, permits decoupling of the gas and dust temperatures in the low density and metallicity regimes.

The mean intensity used in Eqs.~(\ref{eq:contQ}) and (\ref{eq:coll}) accounts for the incident radiation flux on the disk surface and is determined from a simple diffusion approximation that self-consistently accounts for optically thick and thin disk regions \citep{Tanaka2014,Pav2023}
\begin{equation}
    J = {B(T_{\rm irr}) \over 1 + \tau_{\rm P}^2 + 2/3 \tau_{\rm P} \tau_{\rm R} },
\end{equation}
where $\tau_{\rm P}$ and $\tau_{\rm R}$ are the mean Plank and Rosseland optical depths to the disk midplane. The irradiation temperature includes contributions from  both the accretion and photospheric luminosities of the star and also background irradiation
\begin{equation}
    T_{\rm irr}^4 = T_{\rm bg}^4 + {L_\ast \over 4 \pi r^2 \sigma} \cos{\gamma_{\rm irr}}.
\end{equation}
Here, $\sigma$ is the Stephan-Boltzmann constant, $L_\ast$ is the stellar luminosity, and $\cos{\gamma_{\rm irr}}$ is the incidence angle of radiation arriving at the disk surface (with respect to the normal) at radial distance $r$. The background temperature $T_{\rm bg}$ is equal to the initial temperature of the cloud. 
The background temperature is equal to the initial temperature of the cloud.

For the metal fine-structure cooling term ($Q_{\rm metal}$), we do not solve the full non-equilibrium C/O chemistry in the present disk simulations. We adopt the simplified thermo-chemical prescription used in \citet{VorobyovMatsukoba2020} and \citet{met22}, which is based on the minimum model of \citet{om05}. In this prescription, metal cooling is computed from the fine-structure lines of C$^{+}$ ions and O atoms. 
The adopted solar-neighborhood abundances are $y({\rm CII})=9.27\times10^{-5}$ and $y({\rm OI})=3.57\times10^{-4}$ relative to hydrogen nuclei, and these values are scaled down linearly to the metallicity of our fiducial model $Z=0.08 \, Z_\odot$.
The molecular line cooling term includes only H$_{2}$ and HD line cooling. Molecular line cooling by CO and H$_{2}$O is not included because it is important mainly at densities lower than those in the disk regions studied here.

We also evolved a time-dependent non-equilibrium chemical network with 27 reactions to determine mass fractions for the eight species (H, H$_2$, H$^+$, H$^-$, D, HD, D$^+$, and e$^-$) that govern primordial heating and cooling and contribute to molecular line emission. Our network includes the main H$_{2}$ formation channels: the H$^{-}$ channel, formation on dust grains, and three-body reactions. Updates to the energy equation due to $\Lambda$ are performed with a Newton-Raphson root-finding method, complemented with bisection when Newton-Raphson iterations diverge. The other terms in Equation~\ref{eq:energ} are evolved with a time-explicit scheme.  More details on disk cooling and heating can be found in \citet{VorobyovMatsukoba2020,met22}.




\subsection{Dust Dynamics / Growth}

\label{Sect:dustgrowth}

The dust in our model is partitioned into two populations: (i) small dust, which are grains with radii $a_{\rm min}=5\times 10^{-3} \ \mu \rm m$  to $a_{*} = 1 \ \mu \rm m$ and (ii) grown dust from $a_{*}$ to a maximum value $a_{\rm max}$, which is variable in space and time.   Dust in both populations is distributed in radius according to a simple power law $N(a) = Ca^{ - {q}}$, where $N(a)$ is the number of dust particles per unit dust radius, $C$ is a normalization constant, and the index $q = 3.5$, which is kept constant during the evolution of the disk for simplicity.  We evolve separate continuity equations for the grown and small dust ensembles. However, the momentum equation is solved only for the grown dust, because small dust is assumed to be dynamically linked to the gas. The hydrodynamics equations for the two-population dust ensemble in the zero-pressure limit are
\begin{equation}
\label{contDsmall}
\frac{{\partial \Sigma_{\rm d,sm} }}{{\partial t}}  + \nabla  \cdot 
\left( \Sigma_{\rm d,sm} {\bl v} \right) = - S(a_{\rm max}),  
\end{equation}
\begin{equation}
\label{contDlarge}
\frac{{\partial \Sigma_{\rm d,gr} }}
{{\partial t}}  + \nabla  \cdot 
\left( \Sigma_{\rm d,gr} {\bl u} \right) =  
S(a_{\rm max}),
\end{equation}
\begin{equation}
\label{eq:momDlarge}
\frac{\partial}{\partial t} \left( \Sigma_{\rm d,gr} {\bl u} \right) +  \nabla \cdot \left( \Sigma_{\rm d,gr} {\bl u} \otimes {\bl u} \right)  =   \Sigma_{\rm d,gr} \, {\bl g}+ \Sigma_{\rm d,gr} \, {\bl f} + S(a_{\rm max}) {\bl v},
\end{equation}
where $\Sigma_{\rm d,sm}$ and $\Sigma_{\rm d,gr}$ are the surface densities of small and grown dust, respectively, and $\bl u$ are the planar components of the grown dust velocity. Gas and dust are dynamically coupled not only by gravity but also by friction $\bl f$, which is calculated with the Henderson friction coefficient \citep{Henderson1976} as tested in astrophysical contexts in \citet{Stoyanovskaya2020} and implementated in FEOSAD in \citet{VorobyovElbakyan2023}.  This approach allows us to treat drag forces while avoiding discontinuities between the Epstein (free molecular flow) and Stokes (continuous-medium flow) regimes. In the former, the mean free path of hydrogen molecules $\lambda_{\rm m.f.p} \gtrsim a_{\rm max}$, while in the latter $\lambda_{\rm m.f.p} \lesssim a_{\rm max}$ (see Figure~1 in \citealt{Stoyanovskaya2020}).

Small dust can turn into grown dust as the disk forms and evolves. FEOSAD accounts for dust growth in the framework of the monodisperse dust growth model \citep{2012Birnstiel}.  The term $S(a_{\rm max})$ that enters the equations for the dust component is the conversion rate between small and grown dust populations, defined as $S(a_{\rm max}) = - \Delta \Sigma_{\rm d,sm} / \Delta t$, where
\begin{equation}
\label{final}
    \Delta\Sigma_{\mathrm{d,sm}} = \Sigma_{\mathrm{d,sm}}^{n+1}- \Sigma_{\mathrm{d,sm}}^{n} =
    \frac
    {
    \Sigma_{\rm d,gr}^n \int_{a_{\rm min}}^{a_*} a^{3-\mathrm{q}}{\rm d}a - 
    \Sigma_{\rm d,sm}^n \int_{a_*}^{a_{\mathrm{max}}^{\rm n+1}} a^{3-\mathrm{q}}{\rm d}a
    }
    {
    \int_{a_{\rm min}}^{a_{\mathrm{max}}^{n+1}} a^{3-\mathrm{q}}{\rm d}a
    }.
\end{equation}
Here, indices $n$ and $n+1$ denote the current and next hydrodynamic times, respectively, and $\Delta t$ is the hydrodynamic time step. Our dust growth scheme is constructed so as to preserve the total dust mass in each computational cell and also continuity at $a_{*}$ by writing the conversion rate of small to grown dust as $a_{\rm max}$ increases. This scheme effectively assumes that dust growth smooths out any discontinuity in the dust radius distribution at $a_\ast$ that may appear due to differential drift of small and grown dust populations.  This scheme is discussed in greater detail in \citet{VorobyovSkliarevskii2022}.


To find $S(a_{\rm max})$, the time evolution of $a_{\rm max}$ in each computational cell due to collisional growth and advection of dust has to be calculated. The equation describing the dynamical evolution of $a_{\rm max}$ is 
\begin{equation}
\frac{\partial a_{\rm max}}{\partial t} + ({\bl u}_{p} \cdot \nabla_p ) a_{\rm max} = \cal{D},
\label{eq:dustA}
\end{equation}
where the rate of dust growth due to collisions and coagulation is computed in the monodisperse approximation \citep{2012Birnstiel},
\begin{equation}
\cal{D} = \frac{\rho_{\rm d} \mathit{u}_{\rm rel}}{\rho_{\rm s}}.
\end{equation}
This rate includes the total volume density of dust $\rho_{\rm d}$, the dust material  density $\rho_{\rm s} = 3.0$\:g\:cm$^{-3}$, and the relative velocity of particle-particle collisions defined as $\mathit{u}_{\rm rel} = (\mathit{u}_{\rm th}^2 + \mathit{u}_{\rm turb}^2)^{1/2}$, where $\mathit{u}_{\rm th}$ and $\mathit{u}_{\rm turb}$ account for the Brownian and turbulence-induced local motion, respectively. In particular, the dust-to-dust collision velocity due to turbulence is obtained from the turbulent eddy model in \citet{Ormel2007},
\begin{equation}
    v_{\rm{turb}} = \sqrt{{\frac{3 \alpha_{\rm visc}}{\mathrm{St}+\mathrm{St}^{-1}}}} c_{\rm s},
    \label{eq:turb_vel}
\end{equation}
where $c_{\rm s}$ is the sound speed and $\mathrm{St}$ is the Stokes number defined as $\mathrm{St}=t_{\rm stop} \Omega_{\rm K}$. Here, $\Omega_K$ is the Keplerian frequency and $t_{\rm stop}$ is the stopping time for grains with $a_{\rm max}$. The expression for $t_{\rm stop}$ using the Henderson friction coefficient can be found in \citet{VorobyovElbakyan2023}.  Equation~\ref{eq:turb_vel} correctly reproduces a drop in $v_{\rm turb}$ at large $\mathrm{St}$.  When calculating the volume density of dust, we account for dust settling by determining the effective scale height of grown dust $H_{\rm d}$ from the corresponding gas scale height $H_{\rm g}$, $\alpha_{\rm visc}$ parameter, and the Stokes number:
\begin{equation}
    H_{\rm d} = H_{\rm g} \sqrt{\frac{\alpha_{\rm visc}}{\alpha_{\rm visc} + \mathrm{St}}}.
    \label{eq:dust-scale-height}
\end{equation}
%


The dust growth is limited by the fragmentation barrier, defined as the maximum radius of dust grains $a_{\rm frag}$ that can be attained via mutual collisions of dust particles, and by the drift barrier, which is treated self-consistently when solving the dynamical equations for the dust component. The fragmentation velocity is set equal to 5.0~m~s$^{-1}$.  Laboratory experiments provide a range of values from $\sim 1.0$~m~s$^{-1}$ for bare silicates to $\ge10$~m~s$^{-1}$ for icy grains \citep{2008ARA&A..46...21B}.  However, recent data suggest that silicates may be stickier than usual at higher temperatures. In particular, the fragmentation velocity may increase manyfold above 1000~K \citep{2021A&A...652A.106P}.  Since the main focus of our study is on the inner several au where temperatures approach this regime, we have chosen $v_{\rm frag}=5.0$~m~s$^{-1}$.  Effects on total planetesimal mass due to variations in $v_{\rm frag}$ are further discussed in Section~\ref{Sect:paramspace}.

Dust temperatures in the innermost regions of the disk can exceed the threshold for evaporation. To account for this effect we reduced the dust density exponentially in regions where it is sublimated.  The threshold temperature is defined as 
\begin{equation}
T_{\rm d,sub} = 1900 \, [\mathrm{K}] \times \rho_{\rm g}^{0.0195}, 
\label{eq:Tcrit}
\end{equation}
where $\rho_{\rm g}$ is the volume density (in g~cm$^{-3}$) obtained from the corresponding gas surface density and the gas disk scale height $H_{\rm g}$ under the assumption of local hydrostatic equilibrium. With this definition of $T_{\rm d, sub}$, most dust evaporates above 1200~K, in agreement with the ggChem model of dust sublimation and condensation \citep{2018A&A...614A...1W}.  Once evaporated, the dust is removed from the system without being reconstituted at later times.

\subsection{Integration Scheme, Sink Particles and Protostellar Evolution}

\label{App:solution}

The hydrodynamic equations for gas and dust are integrated with the method of finite volumes \citep{SN1992} with dust treated as a pressureless fluid \citep{VorobyovSkliarevskii2022}, and gas and dust are advected with third-order-accurate piecewise-parabolic interpolation \citep{Colella1984}, which ensures low numerical diffusion.   Gas and dust velocity updates due to mutual friction (including backreaction of dust on gas) are performed with a fully implicit scheme that was extensively tested \citep{2018Stoyanovskaya}.  To ease time-step limitations imposed by the Courant condition, a sink cell with a radius of 0.3~au was excised from the center of the grid.  We adopt free inflow-outflow boundary conditions (BCs) at the innermost grid cell, which allows material to flow from the mesh into the sink and vice versa. These BCs are discussed in greater detail in \citet{VorobyovAkimkin2018}, and they perform better than standard outflow BCs because they allow for a compensating flow from the sink to the mesh when oscillatory motions are present at the innermost zone \citep{Zhu2012}.  The effects of BCs on final planetesimal masses are discussed in Section~\ref{App:BCs}.

Matter passing into the sink is assumed to land on the protostar, which gains mass accordingly.  The photospheric luminosity of the protostar is determined with precomputed stellar evolution tracks from the STELLAR code at the given metallicity \citep{Hosokawa2009} using the time-dependent stellar age and mass.  Photospheric and accretion luminosities affect the thermal balance of the disk via stellar irradiation.  Matching the metallicity of the protostar to that of the gas from which it forms is important because photospheric temperatures and luminosities are sensitive to the internal opacities of the star, with lower metallicities producing hotter and more compact objects that are more efficient at evaporating dust near the center of the disk.  We performed tests that showed that assuming higher metallicities for the protostar than for the disk led to significant overestimates of  dust masses, and thus final planetesimal masses. Time-dependent stellar evolution calculations coupled to the disk evolution were not employed in this study.  Possible variations in the internal photospheric luminosity that arise in fully coupled simulations due to variable protostellar accretion are expected to affect the  temperature in the outer, irradiation-dominated regions of the disk, but are of lesser significance in the inner optically thick regions. 


\subsection{Planetesimal Formation}

\label{App:plt}

Dust is converted to planetesimals under the assumption that the streaming instability is the primary mechanism for planetesimal production in protoplanetary disks \citep{Youdin2005}.  Resolution limits and the use of vertically integrated disk quantities make it difficult for FEOSAD to directly model the development of the streaming instability, so we adopt criteria derived from numerical studies of its growth to determine where planetesimals form in the disk \citep{Yang2017}:
\begin{eqnarray}
\label{eq:SI_cond_1}
\log{\xi_{\rm d2g}} \geqslant 0.10 \left(\log{\mathrm{St}}\right)^2 + 0.20 \log{\mathrm{St}} - 1.76    \ \ \ \  (\mathrm{St} < 0.1), \\
\log{\xi_{\rm d2g}} \geqslant 0.30 \left( \log{\mathrm{St}}\right )^2 + 0.59 \log{\mathrm{St}} - 1.57   \ \ \ \  (\mathrm{St} > 0.1),
\label{eq:SI_cond_2}
\end{eqnarray}
where $\xi_{\rm d2g}=\Sigma_{\rm d} / \Sigma_{\rm g}$ is the dust-to-gas mass ratio and $\mathrm{St}$ is the Stokes number.  We also considered the more recent instability criteria from \citet{LiYoudin2021}. These conditions are complemented by the requirement that the volume density of dust in the disk midplane, $\rho_{\rm d}$, be equal to or greater than that of the gas, $\zeta={\rho_{\rm d} / \rho_{\rm g}} \ge 1.0$ \citep{Youdin2005}.  We also require that dust grains be larger than 5.0~mm, following high resolution simulations that suggest that grains larger than 1.0~mm are needed to produce planetesimals via the streaming instability \citep{Carrera2022}.  The value of $\rho_{\rm d}$ is obtained from the surface density of dust $\Sigma_{\rm d,gr}$ via the dust disk scale height $H_{\rm d}$, which scales with turbulent $\alpha_{\rm visc}$ and $\mathrm{St}$ as shown in Equation~\ref{eq:dust-scale-height}.  We note that the use of sub-grid models for the streaming instability is a widespread practice because of difficulty with resolving the process in long-term global disk simulations \citep{Drazkovska2016,Lau2022}.

\begin{figure*}
\centering
\includegraphics[scale=0.6]{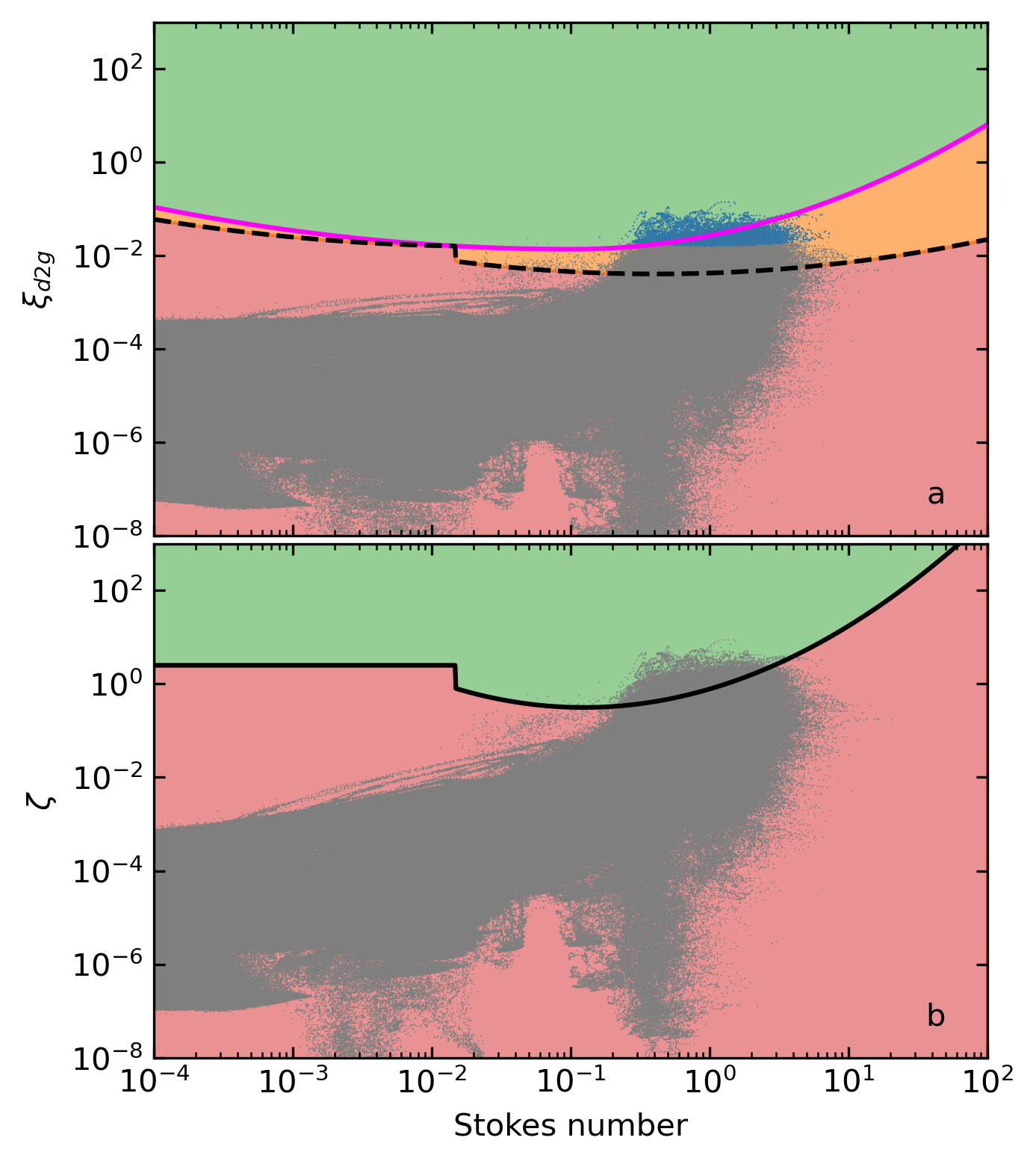} 
\caption{Streaming instability phase space.  Panel (a) shows the ratio of the surface densities of grown dust to gas, $\xi_{d2g}$, as a function of the Stokes number.  The pink and dashed black lines mark the critical values above which the streaming instability develops according to \citet{Yang2017} and \citet{LiYoudin2021}, respectively.  Panel (b) shows the ratio of the volume densities of grown dust to gas, $\zeta$, as a function of Stokes number.  The solid black line indicates the critical values for the streaming instability according to \citet{LiYoudin2021}. The data of our model are overlaid with filled circles, with blue ones fulfilling in addition the criterion on the ratio of volume densities.}
\label{fig:phase2}
\end{figure*}

Dust is converted to planetesimals in regions satisfying these criteria with an approach similar to that in \citet{Drazkovska2016}.  The conversion rate is defined as
\begin{equation}
\label{eq:dpltsm}
    \dot{\Sigma}_{\rm plts} = \chi \, \Sigma_{\rm d,gr}(\mathrm{SI}) \, \psi,
\end{equation}
where $\Sigma_{\rm d,gr}(\mathrm{SI})$ is the surface density of grown dust in regions prone to the streaming instability and $\chi$ is the efficiency of dust to planetesimal conversion, 
\begin{equation}
    \chi = \dfrac{\Omega_{\rm K}}{2\pi \, F},
\label{eq:rate}
\end{equation}
where  $F$ is the characteristic number of local orbital periods that are required to form planetesimals, set here equal to 500 in accordance with \citet{Yang2017}. According to \citet{Carrera2022}, dust grain radii should be substantially larger than 1~mm for the streaming instability to develop in dead zones. In Equation~\eqref{eq:dpltsm} $\psi$ accounts for 
this effect by considering only the fraction of the total dust mass that exceeds the critical threshold dust radius, $a_{\rm thr} = 1.0$~cm:
\begin{equation}
    \psi = \dfrac{\int ^{a_{\rm max}} _{a_{\rm thr}} a^{3-p} da} { \int ^{a_{\rm max}} _{a_{*}} a^{3-p} da}.
    \label{eq:psi}
\end{equation}
We note that $\psi$ is set to zero if $a_{\rm max}<a_{\rm thr}$, thus terminating dust to planetesimal conversion.  Figure~\ref{fig:phase2} shows that the streaming instability can develop in our model disk  according to the criteria laid out in Equations~\ref{eq:SI_cond_1} and \ref{eq:SI_cond_2} and in \citet{LiYoudin2021}. 

\subsection{Boundary Conditions and Gravitational Potential Calculations}
\label{App:BCs}

\begin{figure*}
\centering
\includegraphics[scale=0.6]{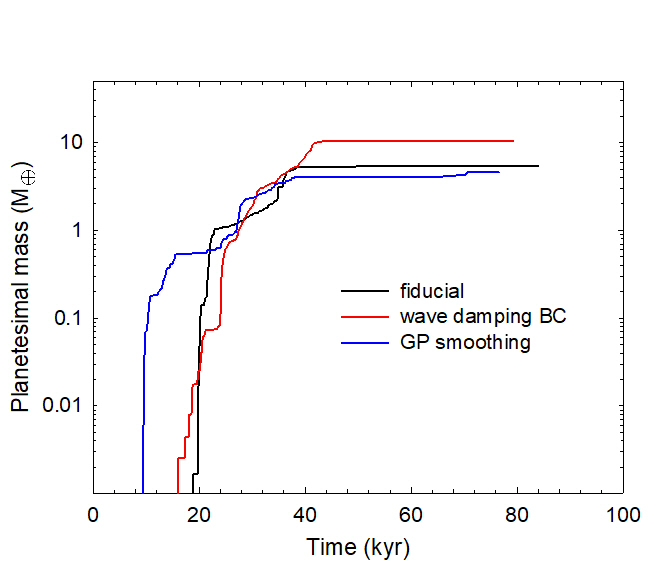} 
\caption{Total planetesimal masses for the fiducial model (black), the model with the wave damping BCs (red), and the model with explicit smoothing of the disk gravitational potential (blue).}
\label{fig:BCs}
\end{figure*}

We show the effect of inner BCs  on the mass of planetesimals $M_{\rm plt}$ in  Figure~\ref{fig:BCs}.  In one test, we adopted the wave damping boundary condition at the sink-disk interface \citep{Zhu2012} instead of the inflow-outflow boundary condition of FEOSAD by allowing matter to flow only from the disk to the sink and limiting this outflow velocity to be no more than 10 times the radial velocity in a steady-state viscous disk.  In this model, $M_{\rm plt}$ reached $\sim$ 10 $M_{\oplus}$, twice that of the fiducial case.  In another test, we introduced explicit smoothing to the disk gravitational potential \citep{Kley2012} instead of the implicit smoothing in FEOSAD \citep{Vorobyov2024}, using $\epsilon=1.2 H$ for the value of explicit smoother \citep{Kley2012}, where the local gas disk scale height $H=0.05 r$.  $M_{\rm plt}$ only fell by a factor of 1.3.  Consequently, the formulation of the inner BCs and disk gravitational potential do not affect our conclusions.

\end{document}